# Reversible Video Steganography Using Quick Response Codes and Modified ElGamal Cryptosystem

**Ramadhan J. Mstafa**[1,2,*]

[1]Department of Computer Science, Faculty of Science, University of Zakho, Duhok, 42002, Kurdistan Region, Iraq
[2]Department of Computer Science, College of Science, Nawroz University, Duhok, 42001, Kurdistan Region, Iraq
*Corresponding Author: Ramadhan J. Mstafa. Email: ramadhan.mstafa@uoz.edu.krd

**Abstract:** The rapid transmission of multimedia information has been achieved mainly by recent advancements in the Internet's speed and information technology. In spite of this, advancements in technology have resulted in breaches of privacy and data security. When it comes to protecting private information in today's Internet era, digital steganography is vital. Many academics are interested in digital video because it has a great capability for concealing important data. There have been a vast number of video steganography solutions developed lately to guard against the theft of confidential data. The visual imperceptibility, robustness, and embedding capacity of these approaches are all challenges that must be addressed. In this paper, a novel solution to reversible video steganography based on Discrete Wavelet Transform (DWT) and Quick Response (QR) codes is proposed to address these concerns. In order to increase the security level of the suggested method, an enhanced ElGamal cryptosystem has also been proposed. Prior to the embedding stage, the suggested method uses the modified ElGamal algorithm to encrypt secret QR codes. Concurrently, it applies two-dimensional DWT on the Y-component of each video frame resulting in Approximation (LL), Horizontal (LH), Vertical (HL), and Diagonal (HH) sub-bands. Then, the encrypted Low (L), Medium (M), Quantile (Q), and High (H) QR codes are embedded into the HL sub-band, HH sub-band, U-component, and V-component of video frames, respectively, using the Least Significant Bit (LSB) technique. As a consequence of extensive testing of the approach, it was shown to be very secure and highly invisible, as well as highly resistant to attacks from Salt & Pepper, Gaussian, Poisson, and Speckle noises, which has an average Structural Similarity Index (SSIM) of more than 0.91. Aside from visual imperceptibility, the suggested method exceeds current methods in terms of Peak Signal-to-Noise Ratio (PSNR) average of 52.143 dB, and embedding capacity 1 bpp.

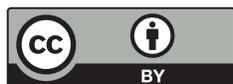

**1 Introduction**

During the past several years, individuals have been more concerned about their data being stolen via the Internet, such as through pirate monitoring and copyright protection, as well as authenticity identification of digital works and identity authentication. Steganography and cryptography have been developed to solve these concerns. Steganography is a science that uses video, image, audio, or other digital media as a medium and then uses a specific algorithm to hide secret data inside the medium. Cryptography, on the other hand, is a science that transforms a secret communication into a meaningless form that eavesdroppers cannot decipher [1–4]. Despite the fact that both steganography and cryptography aim to safeguard data, using either one alone is not the best option. As a result, it is occasionally advised that both techniques be combined. In such a situation, even if the attacker had concerns about the existence of the communication and succeeded to overcome the steganography technique, the perpetrator would still have to crack the encrypted message to acquire the hidden message [5–8].

Steganography algorithms' efficiency is determined by four key aspects: robustness, embedding capability, security, and invisibility. They should be taken into account while creating a new algorithm or when upgrading current ones. The robustness of a steganography algorithm relates to its resilience to threats as well as signal processing. The volume of data that could be embedded inside the cover media is referred to as embedding capacity. The difficulty of an intruder to retrieve the encoded data is referred to as security. The degree of deformation in the original cover carrier due to the concealing procedure is referred to as invisibility [9–11].

Digital video has higher redundancy than other digital media, allowing for a huge amount of data concealing ability. A large number of high-definition videos is also transmitted over the Internet with the big data revolution. As a result, many academics have been interested in video steganography, and it has become a popular option [12,13]. The practice of inserting a private message into a cover video is known as video steganography. It is utilized in a variety of applications, including copyright, remote access, health fields, and enforcement agencies [14–16].

In general, video steganography techniques are classified into three types: format-based approaches, video codec-based approaches, and still image-based approaches [5,17,18]. Still image-based approaches divide a video stream into frames and then implement image steganography techniques to the selected frames to hide data. These approaches are further classified into two types: transform domain techniques and spatial domain techniques. The cover data is first converted into the frequency domain using transform domain tools. The secret message is then used to substitute certain frequency domain values. Lastly, the domain is transformed back into the time domain using the modified values. The transform domain approaches include the DWT, Discrete Cosine Transform (DCT), and Discrete Fourier Transform (DFT). Spatial domain techniques, on the other hand, immediately conceal the sensitive message within the cover carrier data. Because of their low computing cost and easiness, LSB approaches are the most common spatial domain approaches [5,19–21]. LSB techniques work by replacing the sensitive message bits with certain LSBs from the video frames. Videos that use format-based approaches fall into the second type of video steganography. By utilizing the compression approach and structure of a certain video format, these strategies are tailored for a particular video format. Examples of format-based methods include H.264/AVC [22]. In the third category of video steganography, we have approaches that use video codecs as their foundation. They try to utilize the 3D feature of videos and invest the third dimension of insertion, which is the time dimension $t$. In addition to motion components and motion vectors [21,23], this added dimension gives several additional characteristics.

The rest of the paper is structured as follows. Section 2 outlines some state-of-the-art methods related to video steganography. Section 3 explains the types of quick response codes. Section 4 an overview of two-dimensional discrete wavelet transform. Section 5 explains the classical ElGamal cryptosystem. Section 6 describes the proposed reversible video steganography method in detail. Section 7 presents the experimental results with the discussion. Finally, Section 8 concludes the paper.

The main contributions of the proposed approach to the field of information security are as follows:

- We introduce a modified version of the ElGamal cryptosystem so that the size of the encrypted image (QR codes) remains the same as the original image, which addresses the problem of the expansion rate that occurs within the original ElGamal cryptosystem. Moreover, in our approach, when the original image contains multiple pixels with the same values, their encrypted pixels may have different values, making our system highly secure.
- This study offers a novel technique for concealing data into a digital video that is based on two well-known algorithms, the DWT and LSB methods. The suggested technique uses a modified ElGamal cryptosystem to encrypt different types of QR codes first. Then DWT method is applied to each frame of the cover video's Y (luminance) component. Finally, it uses the LSB method to embed the secret QR codes into each pixel of the chosen sub-bands.

## 2 Related Works

Alajmi et al. [24] proposed a valid QR code steganography for encrypted messages. This steganographic method not only hides the payload but also misleads an opponent by using the container to provide false information. To achieve this, a QR code was utilized as a container. In addition to the payload, QR codes created by this technique can carry their normal message. But only a secret key can unlock the message's contents. As a result, a message can be created without respect to its payload and vice versa. A message is generated that sends false information to the attacker. It is demonstrated that the produced QR code is valid, that is, different from an ordinary QR code, making it appear innocent and less susceptible to an opponent's attack. On top of all that, it saves space and is susceptible to steganalysis methods.

Huang et al. [25] proposed an efficient QR code secret embedding method based on Hamming code. This work investigates the characteristics of QR codes in order to suggest an effective secret concealing method for sensitive data contained within QR codes. At the beginning, the secret message is encoded in the cover QR code using the Hamming algorithm (8, 4). Then, the error correcting capability of the QR code fixes the errors that caused during the hidden embedding phase, and the legitimate marked QR code decreases people's expectations. In comparison to existing schemes, this technique outperforms them in terms of secret payload and embedding efficiency. However, this method is not robust against image compression.

Luo et al. [26] proposed EasyStego, a unique cross-domain steganography system. EasyStego is based on the usage of QR codes as carriers, therefore it is resistant to physical distortions in the complicated physical field. Furthermore, EasyStego offers a high capacity for embeddable secrets and great scalability in a variety of circumstances. EasyStego employs an AES encryption technique to manage the authorization of secret messages, which would be more successful in limiting the chance of sensitive information leaking. Experiments demonstrated that EasyStego is resilient and efficient.

Hajduk et al. [27] proposed Image steganography based on QR code and cryptography. This work focuses on the suggestion of the steganographic image method which is used to insert the encoded secret message into image data by use of QR code. DWT domain is being utilized to embed QR code while Advanced Encryption Standard (AES) encryption method also protects the embedding process. Additionally, common features of QR code have been destroyed by encryption making the approach safer. The purpose of this work is to create a very reliable, highly perceptible image steganographic technique. A unique QR code compression before the embedding phase enhanced the relationship between security and the method's capacity. However, related to the size of the covered data, the embedding capability of this method is insufficient.

In order to enhance the security of commercial activities on the internet and media, a strong double-watermarking system for secret code exchange is being developed by Waleed et al. [28]. This scheme is implemented with a DWT as a first fold and a DCT as a second fold for color images that generates unwanted master and secret parts with the same QR code watermark by visual secret share. The experimental results show that the suggested technique is quite resilient, and that the QR code can be decoded even after various attacks are used. However, this method has a low embedding capacity.

Zhang et al. [29] proposed a visible watermarking scheme for QR code based on reversible Data hiding. In this scheme, the QR code can be decoded and the original image can be recovered reversibly after the QR code has been read successfully. Optimization is accomplished by both using QR code features in encoding and decoding in the visible watermarking phase and using the reversible data concealing time to block, scan and preprocess information. The validity and effectiveness of this method has been shown by experimental results. However, the embedding capacity of this algorithm is limited.

## 3 Quick Response Codes

The Denso-Wave Company in Japan first utilized the QR code in 1994 to track automotive components. A lot of other applications made significant use of the code for different object identification. An example of a QR code is shown in Fig. 1 part (a). A QR code is made up of a matrix of modules, each of which is made up of four pixels by four pixels of either black or white squares. For example, the first QR code has modules of $21 \times 21$, the second has modules of $25 \times 25$, the third has modules of $29 \times 29$, and so on. There are four modules more on either side of every version, thus version 40 is $(21 + 4 \times 39) \times (21 + 4 \times 39)$ or, in other words, $177 \times 177$ modules. Position pattern (or finder pattern), which is used to locate the picture of QR code while scanning using barcode scanners, Fig. 1 part (a) illustrates how a QR code's look is mosaic-like with two concentric squares at each of its three corners. In addition, smaller squares of the same form as the position pattern but with a smaller size of $3 \times 3$ modules are termed alignment patterns in the picture of a QR code version greater than 1. These alignment patterns are utilized for accurate module alignment during QR code scanning. It is worth noting that the number of alignment patterns grows in direct proportion to the size of the version. Fig. 1 part (b) shows further information about the QR code's internal structure [30].

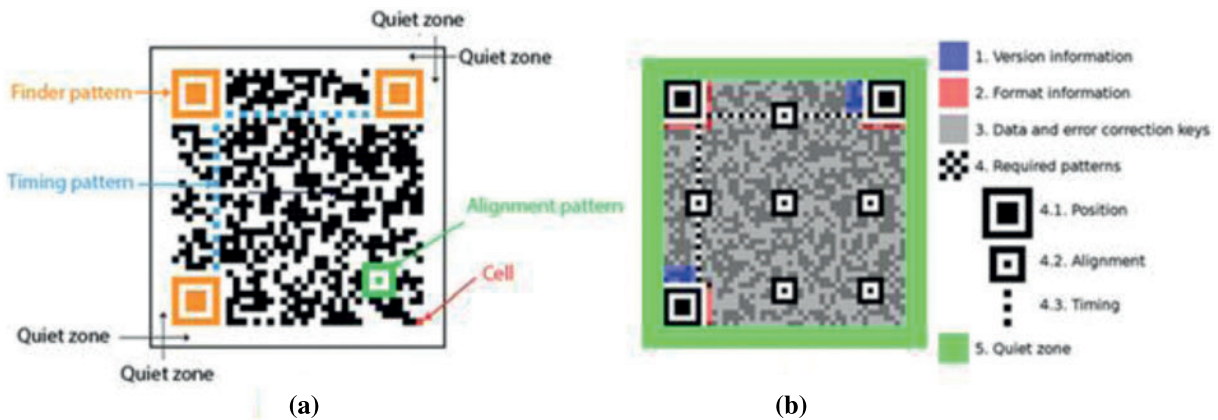

**Figure 1:** QR code. **a)** The QR code structure. **b)** An illustration of the structure for the version-10 QR code, highlighting the functional elements [30]

Each QR code contains 8-bit Reed–Solomon codewords that may be recovered if mistakes occur during the QR code reading process. These codewords are utilized by the Reed–Solomon error correction method and have four degrees of error correction: Low (L), Medium (M), Quantile (Q), and High (H).

QR Code is a matrix-type symbol with a square cell structure. It is made up of the functionality patterns that facilitate reading and the data region where the data is kept. There are finder patterns, alignment patterns, timing patterns, and a quiet zone in QR code [30].

➢ Finder Pattern: A pattern is used to determine the location of the QR code. The location, size, and angle of a sign may be determined by placing this pattern at its three corners. This finder pattern is made up of a structure that may be spotted from all sides (360°).
➢ Alignment Pattern: A pattern for rectifying the QR code's distortion. It is quite good at fixing nonlinear distortions. To repair the symbol's distortion, the central coordinate of the alignment pattern will be discovered. To do this, a black isolated cell is put in the alignment pattern to aid in detecting the alignment pattern's center coordinate.
➢ Timing Pattern: A pattern consisting of black and white patterns is placed alternately to identify the central coordinate of each cell in the QR code. It is used to repair the data cell's central coordinate when the symbol is deformed or when the cell pitch is incorrect. It is placed both vertically and horizontally.
➢ Quiet Zone: A margin area is required to read the QR code. This silent zone facilitates the detection of the sign from among the images received by the charge-coupled device sensor. The quiet zone requires four or more cells.
➢ Data Area: The data from the QR code will be encoded in the data area. Fig. 1's grey region indicates the data area. Based on the encoding rule, the data will be encoded into binary values '0' and '1'. The binary integers '0' and '1' will be translated into black and white cells before being organized. Reed-Solomon codes will be used in the data area to store data and to provide error correcting capabilities.

## 4 Discrete Wavelet Transform

The DWT is a widely used technique, which transforms digital data from the spatial intensities to the transform coefficients. To begin, the two-dimensional DWT is a multi-resolution technique that uses low and high pass decomposition filters to split the video frame into approximation, horizontal, vertical, and diagonal sub-bands. The first level of a two-dimensional DWT decomposition is shown in Fig. 2 with each of the LL, LH, HL, and HH sub-bands. The following wavelet formulas are necessary for complete reconstruction [20]:

$$\{Lo\_D(z) Hi\_D(z) + Lo\_R(z) Hi\_R(z)\} = 2 \quad (1)$$

$$Lo\_R(z) = z^{-k} Hi\_D(-z) \quad (2)$$

$$Hi\_R(z) = z^{k} Lo\_D(-z) \quad (3)$$

where $Lo\_D(z)$ and $Hi\_D(z)$ indicate the decomposition wavelet filters, and $Lo\_R(z)$ and $Hi\_R(z)$ represent the reconstruction wavelet filters. Haar wavelet filters are given in the following questions:

$$Lo\_D(z) = \frac{1}{2}(1 + z^{-1}) \quad (4)$$

$$Hi\_D(z) = (z + 1) \quad (5)$$

$$Hi\_R(z) = \frac{1}{2}(z - 1) \quad (6)$$

$$Lo\_R(z) = (z^{-1} - 1) \quad (7)$$

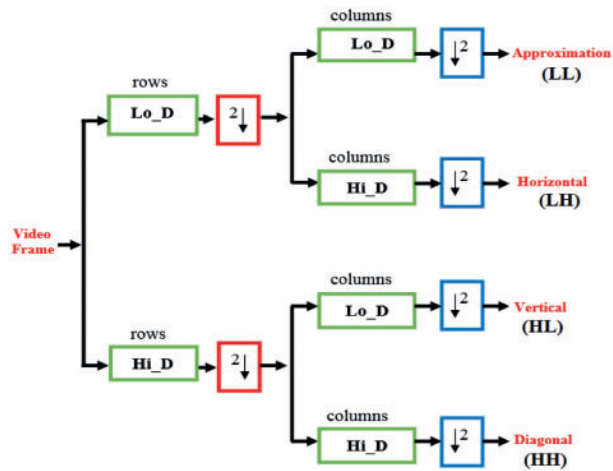

**Figure 2:** First level of a two-dimensional DWT decomposition [20]

## 5 ElGamal Cryptosystem

ElGamal cryptosystem is one of the few non-deterministic schemes, which is considered to be highly secure because, for multiple runs, it generates different cyphertexts for the same plaintext. Its security mainly relies on solving the discrete logarithm problem. This problem occurs when the public key ($p$, $\alpha$, $y$) information is given, and then the private key ($x$) is found, as in Eq. (8). To date, for solving the discrete logarithm problem with a large prime number, there is no known method efficient enough to be feasible for application to cryptanalysis. Thus, it is believed that the ElGamal cryptosystem is extremely secure for a large prime number [31,32].

$$y = \alpha^x \bmod p \tag{8}$$

This cryptosystem includes three stages as follows: key pair generation, message encryption, and message decryption. In the following subsections, these three stages will be elucidated in detail [33].

### 5.1 Key Pair Generation

In this stage, the recipient (User-A) generates the public key and private key information that is essential for the encryption and decryption process. The process for generating these keys is detailed below.

- Generate $p$, a large prime number.
- Select $\alpha$, a primitive root of $p$ such that $1 < \alpha < p - 1$.
- Select $x$, a random integer such that $1 < x < p - 2$.
- Calculate $y$ as follows $y = \alpha^x \bmod p$.

Upon completion of this stage, the recipient (User-A) sends the public key information $(p, \alpha, y)$ to the sender while keeping the private key $(x)$ confidential for later use to decrypt the encrypted message.

### 5.2 Message Encryption

Here, the sender (User-B) uses the public key information $(p, \alpha, y)$ obtained from the recipient (User-A) to encrypt the confidential message. The following steps demonstrate the encryption process in detail.

- Divide the confidential message $m$ into a set of characters.
- Convert each character $m_i$ into its corresponding numerical value, such that $0 \leq m_i \leq p - 1$.
- Select $k$, a random integer such that $1 < k < p - 2$, where $k$ is the private key.
- Calculate $d$ as follows $d = \alpha^k \bmod p$.
- Encrypt each character $m_i$ as follows $z_i = (y^k \times m_i) \bmod p$.

In the end, the sender (User-B) sends the ciphertext information $C = (d, z)$ to the recipient (User-A), where $d$ represents the public key of User-B, and $z$ represents the encrypted message.

### 5.3 Message Decryption

Here, the recipient (User-A) receives the ciphertext information $(d, z)$ from the sender (User-B) to decrypt the encrypted message. A detailed description of the decryption process is given below.

- Calculate $r$ as follows $r = d^{(p-1-x)}$, where $x$ represents the private key of User-A, and $d$ represents the public key of User-B.
- Decrypt each encrypted character $z_i$ as follows $m_i = (r \times z_i) \bmod p$ to obtain the secret message $m$.

## 6 The Proposed Reversible Video Steganography Methodology

In this section, the modified ElGamal cryptosystem is described in depth, followed by the data embedding and extraction processes of the proposed reversible video steganography method:

*6.1 Modified ElGamal Cryptosystem*

It can be noted from the Original ElGamal Cryptosystem (OEC) that the size of the encrypted message ($z$) expands compared to the size of the secret message ($m$), especially when the taken prime number ($p$) is much larger than the maximum character value in the $m$. Another point that can be observed from the OEC is that when the same character appears multiple times in the $m$, it produces the same encrypted character multiple times. For instance, if $m = \textit{'eee'}$, then $z = \textit{'sss'}$. The reason for the latter issue is that the OEC uses this formula $z_i = (y^k \times m_i) \mod p$ to encrypt the entire secret message, where $y^k$ is constant for all characters. To address the above issues, we propose a modified version of the ElGamal cryptosystem (MEC) that uses the same key pair generation algorithm as in OEC but differs in the encryption and decryption process. Moreover, the security of the proposed MEC is analogous to the OEC, which is based on solving the discrete logarithm problem that is difficult to solve at present.

Here, the proposed MEC uses an image as a secret message although it can treat a text as a secret message. Since each pixel in the image occupies one-byte of memory, the proposed MEC randomly generates a set of bytes that have a size roughly equal to the number of pixels in the image. Moreover, to avoid expansion in the image size during the encryption process, it uses an XOR operation between each pixel value and one byte of a randomly generated set of bytes at a time to obtain the encrypted image. By making the above-mentioned modifications to the OEC, the proposed MEC overcomes the limitations of the OEC. Furthermore, it performs better in terms of execution time.

Like other asymmetric cryptosystems, the proposed MEC consists of three phases, namely key pair generation, image encryption, and image decryption. Fig. 3 demonstrates the workflow of the proposed MEC.

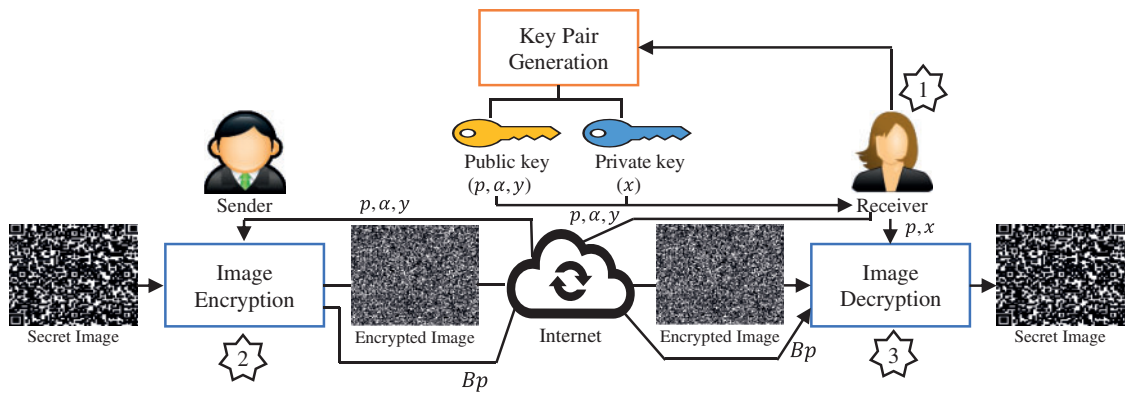

**Figure 3:** A block diagram of the proposed modified ElGamal cryptosystem

*6.1.1 Key Pair Generation*

As mentioned earlier, the key pair generation of the proposed MEC is identical to the OEC. Here, an example is given to show how the public key and private key information are generated.

- Suppose $p = 997, \alpha = 809, x = 420$ are selected randomly, where $p$ represents the prime number, $\alpha$ represents the primitive root of $p$, and $x$ represents the private key.
- Then $y = \alpha^x \mod p = 809^{420} \mod 997 = 12$.

At the end of this stage, the receiver (User-A) sends the public key information ($p = 997$, $\alpha = 809$, $y = 12$) to the sender (User-B) while keeping the private key ($x = 420$) confidential.

### 6.1.2 Image Encryption

Algorithm 1 shows the steps involved in the image encryption process for the proposed MEC. An example is given below to illustrate how the image encryption process works.

Suppose the secret image has a resolution of 3 by 3 as in Fig. 4 part (I), and the public key information received from User-A is $p = 997$, $\alpha = 809$, and $y = 12$. To encrypt the secret image (*im*), the following steps are taken.

**Algorithm 1:** Image encryption algorithm

| | |
|---|---|
| Input: | A secret image (*im*), a public key information ($p$, $\alpha$, $y$) of User-A. |
| Output: | An encrypted image ($z$), a public key information (*Bp*) of User-B. |
| Step1. | Read $p$, $\alpha$, $y$, and *im*. |
| Step2. | Initialize *Bp* and *fsk* as an empty list, for instance, $Bp = []$ and $fsk = []$, where *fsk* represents the final shared secret key. |
| Step3. | Initialize $z$ with zeros so its shape and type are identical to *im*. |
| Step4. | Select $k$, a random integer such that $1 < k < p - 2$, where k represents the private key of User-B. |
| Step5. | Compute $d$ as follows $d = \alpha^k \bmod p$, where $d$ represents the public key of User-B. |
| Step6. | Compute *sk* as follows $sk = y^k \bmod p$, where *sk* represents the shared secret key. |
| Step7. | Append $d$ into *Bp*. |
| Step8. | Convert the *sk* into a set of bytes and then append each byte separately into *fsk*. |
| Step9. | If len(*fsk*) is less than len(*im*) go to step4, where len(*fsk*) returns the number of bytes available inside the *fsk* and len(*im*) returns the size of the *im*. |
| Step10. | The encryption process will be as follows: $z_{i,j} = im_{i,j}\ XOR\ fsk_t$ where: <br> • $i = 0, 1, 2, \ldots, m - 1$, $j = 0, 1, 2, \ldots, n - 1$. <br> • $t = 0, 1, 2, \ldots, r - 1$. <br> • $m$ and $n$ represent the length of row and column in the *im*, respectively. <br> • $r$ represents the length of the *fsk*. <br> • $im_{i,j}$ represents the original pixel value in $i$th and $j$th index of the *im*. <br> • $fsk_t$ represents the shared secret key in $t$th index of the *fsk*. |
| Step11: | $z$ and *Bp* are sent to the recipient (User-A) to decrypt the encrypted image. |

| 12 | 66 | 23 |
|---|---|---|
| 204 | 138 | 76 |
| 0 | 94 | 51 |

I) A secret image (*im*).

| 16 | 65 | 252 |
|---|---|---|
| 205 | 148 | 190 |
| 2 | 15 | 75 |

II) An encrypted image ($z$).

**Figure 4:** The two figures above show the results of obtaining a secret image from a particular encrypted image using the proposed MEC or vice versa

At first, the sender (User-B) has to initialize his public key information and the final shared secret key as follows: $Bp = []$ and $fsk = []$, respectively. After that, he should initialize a new image ($z$) with zeros so that its shape and type are identical to the secret image, where $z$ will be filled with the encrypted pixels during the encryption process. To randomly generate a set of bytes that have a size roughly equal to the number of pixels in the image, Step4 to Step9 from Algorithm 1 are followed. Since we have nine pixels in our example, these steps must generate at least nine bytes. The results of these steps are given below:

- Suppose $k_1 = 87$, $k_2 = 578$, $k_3 = 734$, $k_4 = 55$, $k_5 = 376$, $k_6 = 622$ are selected randomly, where $k$ represents the sender's private key.
- $d_1 = 320$, $d_2 = 619$, $d_3 = 122$, $d_4 = 273$, $d_5 = 171$, $d_6 = 918$, where $d$ represents the sender's public key.
- $sk_1 = 796$, $sk_2 = 491$, $sk_3 = 30$, $sk_4 = 754$, $sk_5 = 81$, $sk_6 = 888$, where $sk$ represents the shared secret key.
- $sk_1 = [28, 3]$, $sk_2 = [235, 1]$, $sk_3 = [30]$, $sk_4 = [242, 2]$, $sk_5 = [81]$, $sk_6 = [120, 3]$. This step converts each shared secret key into a set of bytes. For instance, the first shared secret key ($sk_1 = 796$) is converted as follows: $796_d = 0000\ 0011\ 0001\ 1100_b = [28, 3]$.
- $Bp = [320, 619, 122, 273, 171, 918]$, where $Bp$ represents the public key information of the sender.
- $fsk = [28, 3, 235, 1, 30, 242, 2, 81, 120, 3]$, where $fsk$ represents the final shared secret key that will be used for the encryption process.

To encrypt the secret image given in Fig. 4 part (I), Step10 from Algorithm 1 is followed. The results of this step are given in Fig. 4 part (II). It is important to note here that the size of the $fsk$ in some cases exceeds the size of the secret image as in our example. To avoid this problem, we simply neglect some bytes from the end of the $fsk$ to make them have equal size. For instance, the size of the $fsk$ in our example is ten, and our secret image size is nine. To make them have an identical size, we remove the last byte from the $fsk$ as follows: $fsk = [28, 3, 235, 1, 30, 242, 2, 81, 120]$.

### 6.1.3 Image Decryption

Algorithm 2 shows the steps involved in the image decryption process for the proposed MEC. An example is given below to illustrate how the image decryption process works.

Suppose the encrypted image has a resolution of 3 by 3 as in Fig. 4 part (II), the private key of the receiver (User-A) is $x = 420$, the public key information of User-A is $p = 997$, and the public key information received from User-B is $Bp = [320, 619, 122, 273, 171, 918]$. To decrypt the encrypted image ($z$), the following steps are taken.

**Algorithm 2:** Image decryption algorithm

| | |
|---|---|
| Input: | An encrypted image ($z$), a public key information ($Bp$) of User-B, a public key information ($p$) of User-A, a private key ($x$) of User-A. |
| Output: | A secret image ($im$). |
| Step1. | Read $z$, $Bp$, $p$, $x$. |
| Step2. | Initialize $fsk$ as an empty list, for instance, $fsk = []$, where $fsk$ represents the final shared secret key. |
| Step3. | Initialize $im$ with zeros so its shape and type are identical to $z$. |

(Continued)

| **Algorithm 2:** Continued | |
|---|---|
| Step4. | Divide $Bp$ into a set of blocks as follows: $Bp_1$, $Bp_2$, …, $Bp_{n-1}$, where $n$ represents the length of the $Bp$. |
| Step5. | For each block $Bp_i$, do the following:<br>a) Compute $sk$ as follows $sk = Bp_i^x \bmod p$, where $sk$ represents the shared secret key, and $i$ represents the $i$ th index of the given block.<br>b) Convert the $sk$ into a set of bytes and then append each byte separately into $fsk$. |
| Step6. | The decryption process will be as follows:<br>$im_{i,j} = z_{i,j} \text{ XOR } fsk_t$.<br>where:<br>• $i = 0, 1, 2, …, m-1, j = 0, 1, 2, …, n-1$.<br>• $t = 0, 1, 2, …, r-1$.<br>• $m$ and $n$ represent the length of row and column in the $z$, respectively.<br>• $r$ represents the length of the $fsk$.<br>• $z_{i,j}$ represents the encrypted pixel value in $i$ th and $j$ th index of the $z$.<br>• $fsk_t$ represents the shared secret key in $t$ th index of the $fsk$. |

At first, the receiver (User-A) has to initialize her final shared secret key as follows: $fsk = [\,]$. After that, she should initialize a new image ($im$) with zeros so that its shape and type are identical to the encrypted image, where $im$ will be filled with the secret pixels during the decryption process. To randomly generate a set of bytes that have a size roughly equal to the number of pixels in the image, Step4 to Step5 from Algorithm 2 are followed. The results of these steps are given below.

- $Bp_1 = 320$, $Bp_2 = 619$, $Bp_3 = 122$, $Bp_4 = 273$, $Bp_5 = 171$, $Bp_6 = 918$.
- $sk_1 = 796$, $sk_2 = 491$, $sk_3 = 30$, $sk_4 = 754$, $sk_5 = 81$, $sk_6 = 888$.
- $sk_1 = [28, 3]$, $sk_2 = [235, 1]$, $sk_3 = [30]$, $sk_4 = [242, 2]$, $sk_5 = [81]$, $sk_6 = [120, 3]$.
- $fsk = [28, 3, 235, 1, 30, 242, 2, 81, 120, 3]$.

To decrypt the encrypted image given in Fig. 4 part (II), Step6 from Algorithm 2 is followed. The results of this step are given in Fig. 4 part (I).

### 6.2 Embedding Stage

Data embedding is the process of concealing cryptic information inside cover videos. This procedure divides the streaming video into frames. Each frame converts into the Y, U, and V color space. A unique key ($Key_1$) is used to permute the pixel coordinates of the Y, U, and V components for security reasons. In addition, the QR code is encrypted using a modified ElGamal encryption technique. To reshape the bit locations, $Key_1$ has been used to permute the complete bit positions of the QR message. The embedding procedure is accomplished by performing a two-dimensional discrete wavelet transform (2D-DWT) on the Y-component of each video frame. The encrypted Low QR (LQR), Medium QR (MQR), Quantile QR (QQR), and High QR (HQR) codes are then inserted into the HL, HH, U, and V components, respectively. Therefore, the pixels of the YUV components will be rearranged in relation to the original frame pixel locations in order to create the stego frames. As a result, the inverse two-dimensional discrete wavelet transform (2D-IDWT) will be applied on the Y-component of each video frame. Finally, the stego video is constructed from these stego frames. Fig. 5 shows the block diagrams of the data embedding phase.

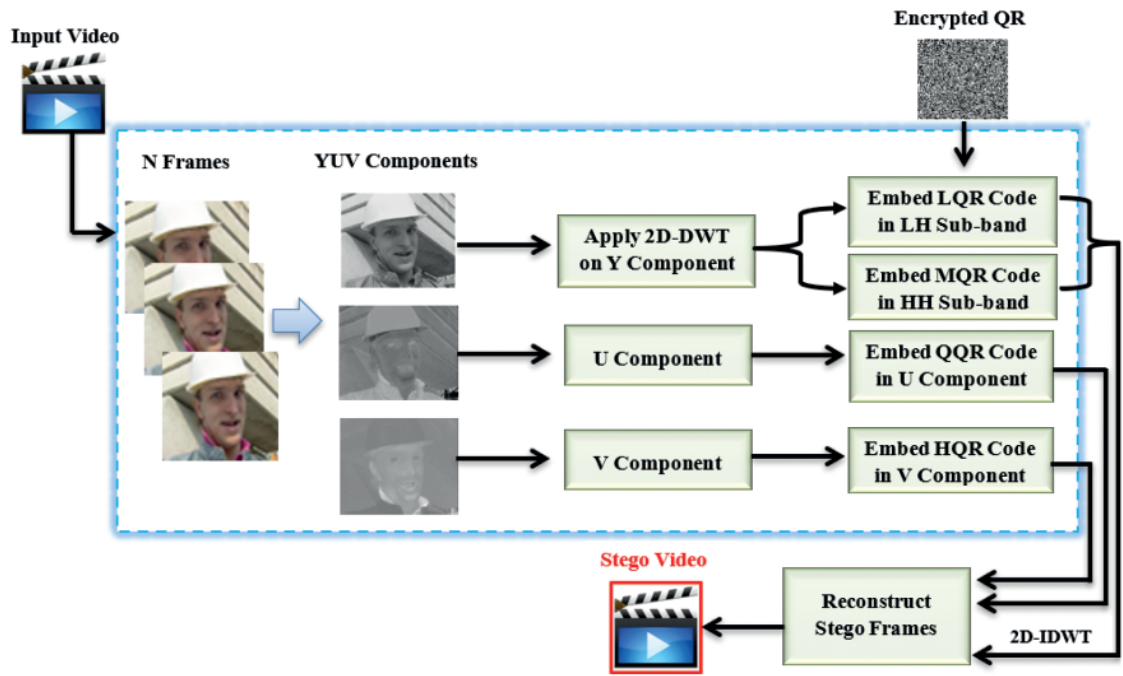

Figure 5: The block diagrams of the data embedding phase

### 6.3 Extracting Stage

The hidden message in the stego videos can be retrieved using a procedure known as data extraction. The embedded videos are turned into frames in order to obtain the exact hidden information. After then, each frame is broken down into its components Y, U, and V. $Key_1$ is used to permute the pixel coordinates in all three Y, U, and V components. By extracting the LSB from each chosen pixel, the secret message can be extracted from YUV components. Each video frame's Y component is subjected to 2D-DWT for extracting purposes. Next, the encrypted LQR, MQR, QQR, and HQR codes will be retrieved for each of the HL, HH, U, and V portions, respectively. Then, each QR code type is decrypted using a modified ElGamal decryption technique. Finally, the permutation operation will be applied on the four types of QR codes to restore their original bit order since all bits of the secret message have been permuted before data embedding process. Fig. 6 shows the block diagrams of the data extraction phase.

## 7 Experimental Results and Discussion

In this section, the suggested method's efficiency is examined in depth. In the first place, we will go through the data collection and assessment metrics that we employed in these experiments. Afterward, the results of the suggested method are discussed. Finally, the suggested approach is compared and widely discussed with existing methods in the literature.

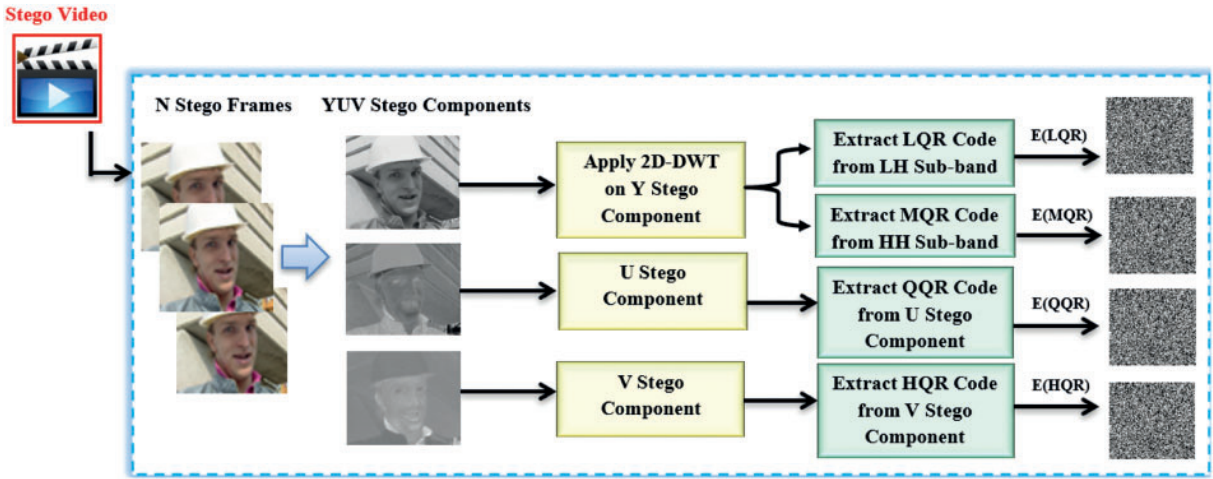

**Figure 6:** The block diagrams of the data extraction phase

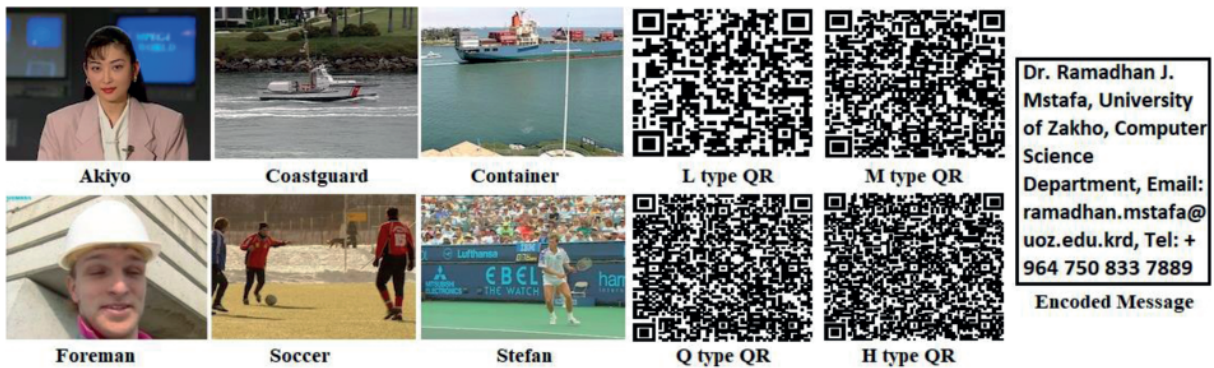

**Figure 7:** Set of test videos and L, M, Q, H QR code types used for the experimental results

### 7.1 Dataset

In order to test the suggested reversible video steganography approach, a dataset of six regularly used video sequences was employed. This dataset was found from the reference [34]. Each of Akiyo, Coastguard, Container, Foreman, Soccer, and Stefan CIF video sequences have been utilized in this work with the resolution of 352 by 288. It was decided that the secret message would be different types (L, M, Q, and H) of QR codes of size 176 by 144 pixels, as it is shown in Fig. 7. A personal PC with the following characteristics was used for our work: MATLAB (R2017b). An Intel Core i7 2nd Generation processor with 8 cores, 2.2 GHz, 6144 MB of DDR3 RAM, and 2034 MB of Radeon 6000 series VRAM.

### 7.2 Evaluation Metrics

Video steganography approaches face the challenge of including as much information as possible into the cover video while maintaining little visual impact on the stego video. As a result, two measures, namely embedding capacity and imperceptibility, were used to assess and compare the suggested technique with existing methods. Capacity is the amount of information that can be concealed in the cover video, which is measured in bits-per-pixel (bpp) and computed as in Eq. (9) [35,36].

$$\text{Embedding Capacity} = \frac{\text{Number of embedded bits}}{\text{Cover video size in pixels}} \times 100\% \text{ (bpp)} \quad (9)$$

The visual quality of the stego videos is used to assess the second parameter, imperceptibility. PSNR, which is measured in decibels (dB) and derived as in Eq. (10) is often used to quantify this statistic. PSNR values below 30 dB show that the human eye is able to detect the distortion, which means that the distortion is noticeable. Steganography algorithms should thus aim for signal-to-noise ratios of at least 40 dB [37,38].

$$\text{PSNR} = 10 \times \log_{10}\left(\frac{\text{MAX}_A^2}{\text{MSE}}\right) \text{(dB)} \quad (10)$$

Mean squared error (MSE) is determined as in Eq. (11).

$$\text{MSE} = \frac{\sum_{i=0}^{a-1} \sum_{j=0}^{b-1} \sum_{k=0}^{c-1} [A(i,j,k) - B(i,j,k)]^2}{a \times b \times c} \quad (11)$$

where $A$ and $B$ represent the original and stego frames, respectively, $a$ and $b$ denote the resolution of the given video, $c$ refers to the number of components of each video frame (for RGB color space, $c = 3$). $MAX_A$ refers the greatest pixel value in frame $A$.

In addition, the robustness measure was utilized to assess the performance of the suggested method against various attacks (such as Salt & Pepper noise, Gaussian noise, Poisson noise, and Speckle noise). This statistic calculates the percentage of similarity between the original QR codes and the QR codes that have been extracted. This metric was measured using the SSIM function, which is defined mathematically as in Eq. (12) [39]. Images with higher similarity scores are more likely to be of a high grade.

$$\text{SSIM} = \frac{(2\mu_O\mu_E + C_1)(2\sigma_{OE} + C_2)}{(\mu_O^2 + \mu_E^2 + C_1)(\sigma_O^2 + \sigma_E^2 + C_2)}$$

where $O$ represents the original QR code, $E$ represents the extracted QR code, $\mu_O$ and $\sigma_O$ represent the mean and standard deviation values of pixels in QR code $O$, respectively, $\mu_E$ and $\sigma_E$ represent the mean and standard deviation values of pixels in QR code $E$, respectively, $C_1$ and $C_2$ refer to a fixed value, $\sigma_{OE}$ represents the covariance between $O$ and $E$ QR codes.

### 7.3 Results of the Proposed Method

This section demonstrates the embedding capacity, PSNR, and SSIM of the proposed technique on six cover videos. Encrypted QR codes of different types were embedded into the LH, HH, U, and V components of each video frame using 2D-DWT. In order to enhance the hidden data in each cover video without negatively compromising the quality of the stego video, we used empirical experiments to determine these areas for the embedding process.

Tab. 1 displays the performance of the suggested method on six cover videos in terms of embedding capacity and PSNR. A greater embedding capacity rate of roughly 1 bpp can be noticed in Tab. 1 for all videos including "Akiyo", "Coastguard", "Container", "Foreman", "Soccer", and "Stefan". An adequate quantity of information may be hidden using the suggested method. According to Tab. 1, the PSNR values for all used videos are larger than 52.139 dB, which is consistent with previous studies. Proposed methods have a high degree of perceptual invisibility. Tab. 1 shows that the overall PSNR averages 52.143 dB among all tested videos. There is no doubt that the suggested approach has a very low level of detection. This means that our technique has a high degree of imperceptibility and an appropriate embedding capacity. In Fig. 8, the suggested method's performance in terms of PSNR is shown for each tested video throughout its 300 frames. As can be seen in Fig. 9, the MSE values of each cover video vary from 0.41 to 0.42, indicating the least amount of degradation in video quality.

Table 1: The embedding capacity and PSNR of the proposed method using six video sequences

| YUV video sequence | Video size in pixels | Total number of embedded bits | Embedding capacity (bpp) | PSNR (dB) |
|---|---|---|---|---|
| *Akiyo* | 30,412,800 | 30,412,800 | 1 | 52.139 |
| *Coastguard* | 30,412,800 | 30,412,800 | 1 | 52.144 |
| *Container* | 30,412,800 | 30,412,800 | 1 | 52.141 |
| *Foreman* | 30,412,800 | 30,412,800 | 1 | 52.142 |
| *Soccer* | 30,412,800 | 30,412,800 | 1 | 52.145 |
| *Stefan* | 30,412,800 | 30,412,800 | 1 | 52.147 |
| Average | | | 1 bpp | 52.143 dB |

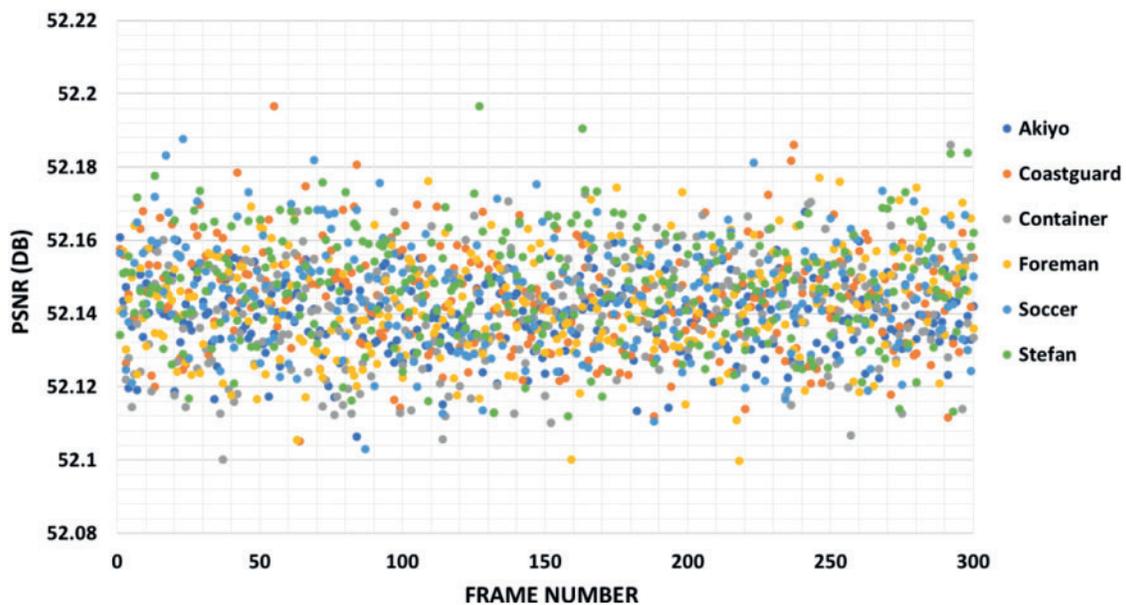

Figure 8: PSNR of the proposed method using six video sequences

Tab. 2 shows how the proposed method performs in terms of SSIM rates on six cover videos with and without noises. Tab. 2 shows that when no noise is introduced to the cover videos, the SSIM rate is "1." This suggests that the confidential information may be retrieved without compromising its security. When the suggested method is used tested videos that have Salt and Pepper noise with a density of 0.01, it still provides a high SSIM rate of around 0.97, which is very near to the SSIM rate of cover videos without noise. However, when the density of Salt and Pepper noise rises, the SSIM rate reduces. When using Gaussian noise in the cover videos, our proposed method achieves SSIM rates of around 0.94 (mean = 0 and variance = 0.01), as shown in Tab. 2. However, the SSIM rate decreases to 0.76 when Gaussian noise has a mean = 0 and variance = 0.1, which is still acceptable. In addition, when Poisson and Speckle noises are incorporated into the cover videos, which results in SSIM rates of more than 0.99 and 0.96, respectively.

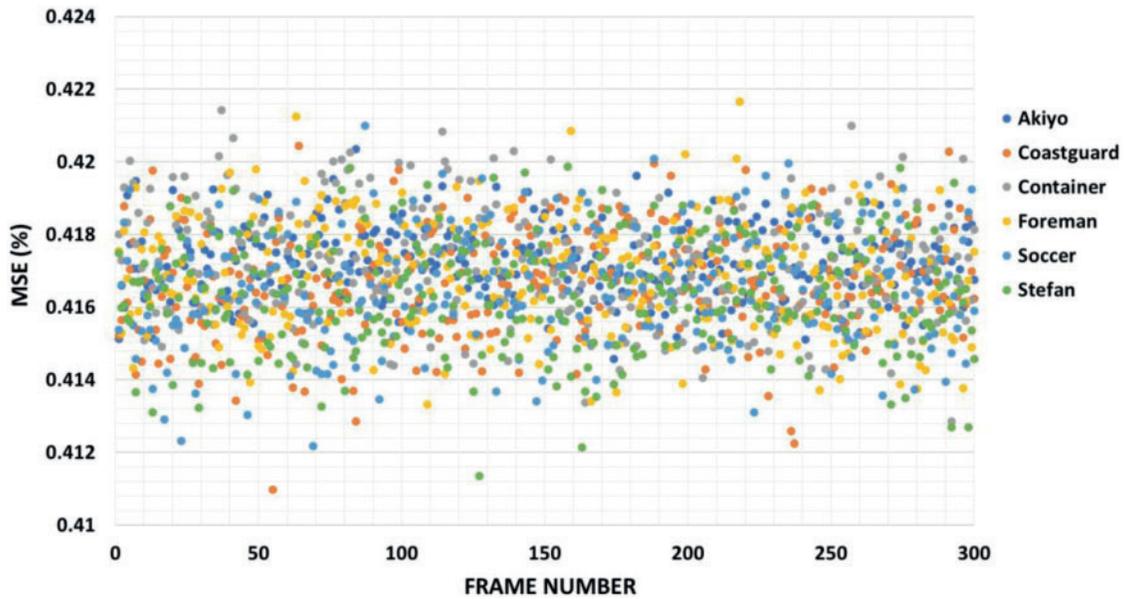

**Figure 9:** MSE of the proposed method using six video sequences

**Table 2:** The performance of the proposed method in terms of SSIM for the recovered L, M, Q, and H QR codes under different types of attack to decode the secret message

| Attacks | SSIM for L QR codes | SSIM for M QR codes | SSIM for Q QR codes | SSIM for H QR codes | Decoded L, M, Q, and H QR codes |
|---|---|---|---|---|---|
| *No attacks* | 1 | 1 | 1 | 1 | Decodable |
| *Salt and Pepper D = 0.01* | 0.9750 | 0.9773 | 0.9838 | 0.9813 | Decodable |
| *Salt and Pepper D = 0.1* | 0.7962 | 0.8135 | 0.8376 | 0.8450 | Decodable |
| *Gaussian (M = 0 and V = 0.01)* | 0.9276 | 0.9452 | 0.9607 | 0.9636 | Decodable |
| *Gaussian (M = 0 and V = 0.1)* | 0.7314 | 0.7636 | 0.7852 | 0.7940 | Decodable |
| *Poisson* | 0.9955 | 0.9974 | 0.9978 | 0.9980 | Decodable |
| *Speckle V = 0.05* | 0.9550 | 0.9658 | 0.9711 | 0.9732 | Decodable |

Using Tab. 2, we can infer that the proposed method is resilient when the cover videos are devoid of noises; nevertheless, the proposed method suffers when noises are included in the cover videos. Salt and pepper noises, on the other hand, are more suited to the suggested method than other noises. Using Poisson and Speckle noises, the suggested approach is more effective than using Gaussian noise in obtaining a higher SSIM rate. Changing frame pixel values have a tendency to be noisy. Fig. 10 illustrates a sample resultant of recovered QR codes under different types of attacks.

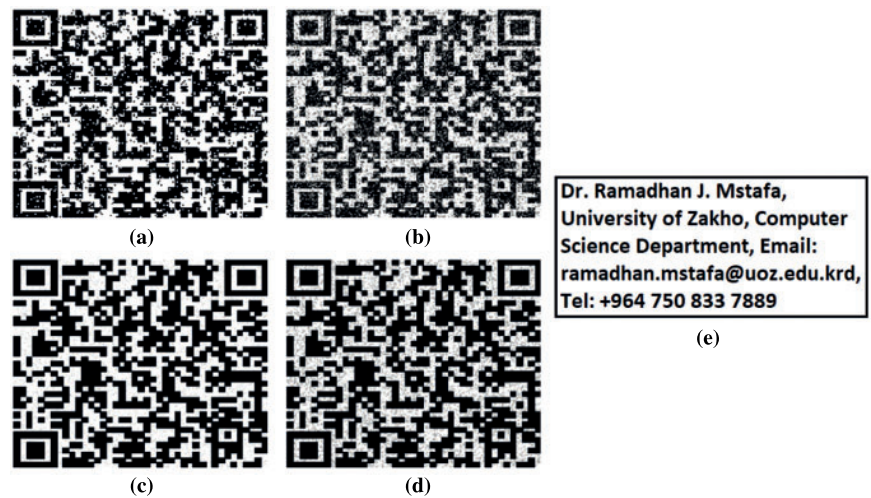

**Figure 10:** The resultant of recovered QR codes under different types of attacks **a)** Recovered L-QR code with SSIM = **0.7962** under Salt and Pepper attack with the density = 0.1 **b)** Recovered L-QR code with SSIM = **0.7314** under Gaussian attack with mean = 0 and variance = 0.1 **c)** Recovered L-QR code with SSIM = **0.9955** under Poisson attack d) Recovered L-QR code with SSIM = **0.9550** under Speckle attack with the variance = 0.05 **e)** The decodable message from L, M, Q, and H QR codes under different types of attacks

### 7.4 Comparisons with Other Approaches

In this section, the perceptual invisibility and embedding capacity of the suggested method were compared with the existing methods from the literature. It was decided to compare PSNR rates of the suggested approach with those of the methods in [36,40–42]. Tab. 3 shows that the suggested technique has the greatest PSNR rate compared to the methods provided in [36,40,41]. Tab. 3 shows that the suggested method's average PSNR rate is superior to the approaches given in [36,40,41]. Compared to the average PSNR rate in [36,40,41], this is an improvement of 3.11, 16.42, and 0.036 dBs. Despite the fact that the technique described in [42] is superior, the proposed method still has a highest total number of embedded bits.

It was decided to compare the suggested approach to the methods described in [36,40–42] in terms of embedded bits. Tab. 3 shows that the suggested technique has the largest overall number of embedded bits compared to the methods provided in [36,40–42] across all videos utilized. Tab. 3 further show that the average number of embedded bits achieved by the suggested technique is substantially larger than the methods described in [36,40–42].

Table 3: Comparison of the proposed method with other existing approaches in terms of embedding capacity and PSNR

| Method | Average of embedding capacity (bits) | Average of PSNR (dB) |
|---|---|---|
| Ref. [36] | 47,081 | 49.033 |
| Ref. [40] | 20,165 | 35.72 |
| Ref. [41] | 7,181,920 | 52.107 |
| Ref. [42] | 1,889,314 | 60.531 |
| *Proposed algorithm* | 30,412,800 | 52.143 |

However, despite the fact that embedding capacity and visual quality are contradictions, the suggested method has managed to achieve an outstanding compromise between the two. A superior visual imperceptibility and embedding capacity may be inferred from the result of this research, compared to approaches described in [36,40–42].

## 8 Conclusions

This paper proposes a reversible video steganography method based on DWT and modified ElGamal cryptosystem using different types of QR codes. The proposed method encrypts the secret information using modified ElGamal algorithm prior to the embedding process to improve the security of the secret data. For hiding the encrypted QR codes, the proposed method applies 2D-DWT on the Y component of each video frame. After that, the proposed method hides the encrypted L, M, Q, and H QR codes into the HL, HH, U, and V components, respectively using LSB algorithm. From the experimental results, it can be seen from the results that the proposed method performs better than the methods presented in [36,40–42] in terms of embedding capacity. In addition, it is clear that the proposed method outperforms state-of-the-art methods presented in [36,40,41] in terms of visual imperceptibility. Although the method presented in [42] outperforms the proposed method in terms of visual imperceptibility, the proposed method is still better in terms of embedding capacity. Furthermore, the given results showed the acceptable range with regard to robustness against different artificial attacks (Salt and Pepper noises with a density of (0.1, and 0.01), Gaussian noises with variance (0.1, and 0.01), Poisson noises, and Speckle noises).

**Funding Statement:** The authors received no specific funding for this study.

**Conflicts of Interest:** The authors declare that they have no conflicts of interest to report regarding the present study.